\documentclass{ws-p10x7}

\def\Journal#1#2#3#4{{#1} {\bf #2}, #3 (#4)}

\def\NIMA{{\em Nucl. Instrum. Methods} A}
\def\NPB{{\em Nucl. Phys.} B}
\def\NPA{{\em Nucl. Phys.} A}

\def\PRD{{\em Phys. Rev.} D}

\begin{document}

\title{BES Recent Results and Future Plans}

\author{Z. G. Zhao \\ 
        Representing the BES Collaboration}

\address{Institute of High Energy Physics, 19 Yuquan Road, Beijing 100039,
China\\E-mail: zhaozg@pony1.ihep.ac.cn}

\twocolumn[\maketitle\abstract{
We report the preliminary $R$ values for all the 85 energy points scanned in
the energy region of 2-5 GeV with the upgraded Beijing Spectrometer (BESII)
at Beijing Electron Positron Collider (BEPC). Preliminary results from the
$J/\psi$ data collected with both BESI and BESII are presented. Measurements 
of the branching fraction of the $\psi(2S)$ decays and the $\psi(2S)$ 
resonance parameters are reported. The future plans, i.e. significantly 
upgrade the machine and detector are also discussed.}] 

\section{Introduction}
The Beijing Spectrometer (BES) at Beijing Electron Positron 
Collider (BEPC), the only operating $e^+e^-$ machine in the world which 
covers the center-of-mass energy in 2-5 GeV and directly produce 
charmonium and charmed mesons, has been running for more than 10 years.  
Both BEPC and BES were upgraded from 1995 to 1997. BEPC has single bunch and
the peak luminosity reached so far is about $5 \times 10^{30} /cm^2 s$ 
at the $J/\psi$ resonance, giving a hadronic event rate about 6-8 Hz.
The Beijing Spectrometer (BES) is a conventional cylindrical magnetic
detector and is described in detail in Ref. 1 and 2.
Table 1 listed some major parameters of the performance of BESII. 

\begin{table}[htp]
\begin{center}
\caption {The major parameters of the performance of BESII}
\vspace{0.2cm}
\begin{tabular}{ccc}
\hline
 System & Parameter           & BESII          \\\hline
 VC     & $\sigma_{xy}$($\mu$)& 100                 \\
 MDC    & $\sigma_{xy}$($\mu$)& 190-210            \\
        & $\Delta p/p$        & 1.78$\%\sqrt{(1+p^2)}$ \\
 TOF    & $\sigma_t$(ps)      & 180                \\
 BSC    & $\sigma_E/E$    & 22$\%/\sqrt{E}$         \\\hline
\end {tabular}
\end{center}
\end{table}

About $7.8 \times 10^6$, $3.96 \times 10^6$ $J/\psi$ 
and $\psi(2S)$ events have been collected with BESI, and $22 \times 10^6$ 
$J/\psi$ events with BESII, in addition to the $R$ scan done with BESII
in the energy region of 2-5 GeV. 
This presentation will firstly report the new measurement of the $R$ values 
in the 2-5 GeV energy region, and then present recent results from the
analysis of the BES $J/\psi$ and $\psi(2S)$ data. All results reported in
this article are preliminary. The future plans for the BES will be briefly 
discussed in the end.   

\section{New $R$ values from the BES}
The QED running coupling constant evaluated at the $Z$ pole,
$\alpha(M^2_{Z})$, and the anomalous
magnetic moment of the muon, $a_{\mu}=(g-2)/2$, are two fundamental
quantities to test the Standard Model(SM)~\cite{blondel,zhao}.
$\alpha(M^2_{Z})$, as
one of the three input parameters in the global fit to the
electroweak data, is sensitive to the predicted mass of the Higgs.
Theoretically, $a_{\mu}$ is sensitive to large energy
scales and very high order radiative corrections~\cite{carey}.
Any deviation between the SM predicted value of anomalous magnetic
moment of the muon, $a_{\mu}^{SM}$, and that from the experimentally
measured one, $a_{\mu}^{exp}$, may hint new physics.  However, the
uncertainties in both $\alpha(M^2_{Z})$ and $a_{\mu}^{SM}$ are
dominated
by the hadronic vacuum polarization, which cannot be reliably
calculated but are related to $R$ values through dispersion
relations~\cite{zhao}. Here $R$ is the lowest order cross section for
$e^+e^-\rightarrow\gamma^*\rightarrow \mbox{hadrons}$, which is
defined as
$R=\sigma(e^+e^- \rightarrow \mbox{hadrons})/\sigma(e^+e^-\rightarrow
\mu^+\mu^-)$,
where the denominator is the lowest-order QED cross section,
$\sigma (e^+e^- \rightarrow \mu^+\mu^-) = \sigma^0_{\mu \mu}=
4\pi \alpha^2 / 3s$.

Since the uncertainties in $\alpha(M^2_{Z})$ and
$a^{SM}_{\mu}$ are limited by the second order loop effects from the 
hadronic vacuum polarization, which dominated by the errors of the
values of $R$ in the cm energy range below 5 GeV~\cite{zhao}, it is
crucial to reduce the uncertainties in the $R$
values measured about 20 years ago with a precision of about
15-20\% in the energy region of 2-5 GeV~\cite{blondel,carey}.

Following the first $R$ scan with 6 energy points in 2.6-5 GeV range
done in 1998~\cite{besr_1}, the BES collaboration performed a finer
$R$ scan with 85 energy points in the energy region of 2-4.8 GeV.
To understand the beam associated background,
separated beam running was done at 24 energy points and single
beam running for both $e^-$ and $e^+$ was done at 7 energy
points distributed over the whole scanned energy region. Special runs
were taken at the $J/\psi$ to determine the trigger efficiency.
The $J/\psi$ and $\psi(2S)$ resonances were scanned at the beginning and
at the end of the $R$ scan for the energy calibration.

The $R$ values from the BESII scan data are measured by observing the
final hadronic events inclusively, i.e. the value of $R$ is determined
from the number of observed hadronic events ($N^{obs}_{had}$) by the
relation
\begin{equation}
R=\frac{ N^{obs}_{had} - N_{bg} - \sum_{l}N_{ll} - N_{\gamma\gamma} }
{ \sigma^0_{\mu\mu} \cdot L \cdot \epsilon_{had}
\cdot \epsilon_{trg}
\cdot (1+\delta)},
\end{equation}
where $N_{bg}$ is the number of beam associated background events;
$\sum_{l}N_{ll},~(l=e,\mu,\tau)$ and $N_{\gamma\gamma}$ are the numbers
of misidentified lepton-pairs from one-photon and two-photon processes
events respectively; $L$ is the integrated luminosity; $\delta$ is
the radiative correction; $\epsilon_{trg}$ and $\epsilon_{had}$ are
trigger and detection efficiency for hadronic events repectively.

The typical errors from different sources at 3.0 GeV are listed in 
table 2.
To further improve the measurement of $R$ values at BEPC, one needs
better detector performance and improve the hadronic event generator, 
as well as higher machine luminosity, particularly for the energies 
below 3.0 GeV.

\begin{table}
\caption{Error Sources for $E_{cm}$=3.0 GeV. Adding the systematic
and statistic errors in quadrature gives a total error of 5.8\%.}
\begin{center}
\begin{tabular}{cccccc} \hline
Source & $N_{had}$ & $L$ & $\epsilon_{had}$ & $1+\delta$ & Stat.\\
Err.(\%) & 3.3 & 2.3 & 3.0 & 1.3 & 2.5\\ \hline
\end{tabular}
\end{center}
\end{table}

\begin{figure}
\centerline{\psfig{figure=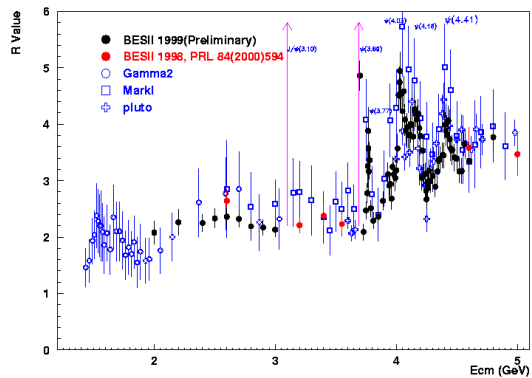,width=60mm,angle=-90}}
\caption{$R$ values below 5 GeV. \label{fig:besr}}
\end{figure}

The preliminary $R$ values obtained at all 85 energy are
graphically displayed in Fig.~\ref{fig:besr}, together with the
6 energy points measured in the first scan and those measured by MarkI,
$\gamma\gamma 2$ and Pluto about twenty years ago.
The preliminary $R$ values from BESII have an average uncertainty of
about 7\%
and are slightly lower than that from the previous measurements. The
two to three factor improvement in precision of the $R$ values in 2-5 GeV
has a significant impact on the global fit to the electroweak data for the
determination of $m_H$. The preliminary fit results show that the
predicted $m_H$ is significant increased with the preferred
value lying just above the LEP2 excluded region, and the new $\chi^2$
profile
of the fit accommodates the LEP2 bound on the mass more
comfortably~\cite{bolek,martin}.
On the other hand, BESII $R$ values is also important to the
interpretation of the E821 $g-2$ measurement~\cite{carey}.

\section{Results on charmonium physics}

The decays of $J/\psi$ and $\psi(2S)$ is an unique laboratory to 
systematically study the charmonium family members, search for gluonic
matters and test of QCD.   

\subsection{Study of the $J/\psi$ decays}
Glueball, the bound state of gluons, has been predicted for all 
low-energy approximations of QCD. Glueball spectrum is one of the least
understood features of the Standard Model. The present consensus is 
that the ground state should be in the 1-2 GeV mass range, and there is
little doubt that radiative $J/\psi$ decays are the best place to search for
glueballs.     

\bigskip
\noindent{\it{3.1.1 Results from BESI $J/\psi$ data}}\\

\medskip
\noindent $\bullet$ Partial Wave Analysis on $J/\psi \to \gamma K^+ K^-$

\medskip
Based on BESI $7.8 \times 10^6$ $J/\psi$ data, a partial
wave analysis is performed to the $f_J(1710)$ mass region in
$J/\psi \to \gamma K^+ K^-$ channel. Fig. 2 shows the mass projection of
the components in the fit to the real data. Apparently, $0^{++}$ is
dominant in $f_J(1710)$ mass region.

\begin{figure}[hbt]
\centerline{\hbox{\psfig{file=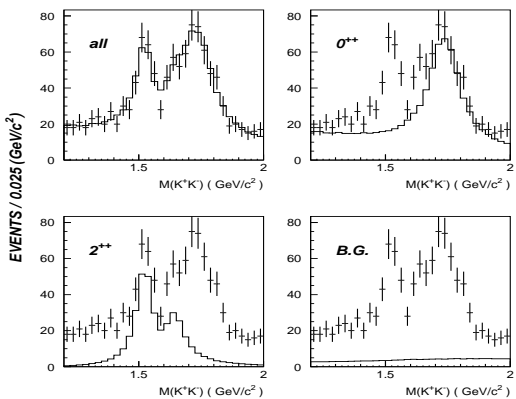,height=60mm,width=75mm}}}
\vspace*{-0.3cm}
\caption{Projection of the different components on the $K^+ K^-$ mass}
\label{fig2:fj_proj}
\end{figure}         

\bigskip
\noindent $\bullet$  The analysis of $J/\psi \rightarrow \gamma \gamma
V(V=\rho,\phi)$

\medskip
We studied the $J/\psi \rightarrow \gamma\gamma V(V=\rho,\phi)$ decay
channels and found the evidence of the $\eta(1430)$ in both decay channels.
The mass and width of the $\eta(1430)$ resonance are found to be $1431 \pm
17~MeV$ and $88 \pm 28~MeV$ from $\gamma \gamma \rho$ mode and $1427 \pm
13~MeV$ and $77 \pm 46~MeV$ from $\gamma \gamma \phi$ mode~\cite{xugf}.
In order to understand the structure of
$\eta(1430)$, other $J/\psi$ decay channels should be analyzed.

\bigskip
\noindent $\bullet$  $N^*$ results from $J/\psi$ decay~\cite{lihb}

\medskip
The behaviour of $N^*$ baryon in $J/\psi$ decay to baryon-antibaryon final
states
provides us a laboratory for the study of $N^*$ baryon, especially in the
mass range of ~1-2 GeV. Using PWA, we have studied the $J/\psi \rightarrow
p\bar{p} \eta$ and $p\bar{p} \pi^0$ decays. Two resonances with
$J^P=\frac{1}{2}^-$ are observed. Their masses and widths are:
$M = 1540^{+15}_{-17}$ MeV with $\Gamma =178^{+20}_{-22}$ MeV; and
$M = 1648^{+18}_{-16}$ MeV with $\Gamma = 150 $ MeV respectively. These two
states are considered to be the nucleon resonances $S_{11}(1535)$ and
$S_{11}(1650)$. 

\bigskip                        
\noindent{\it{3.1.2 Preliminary results from BESII $J/\psi$ data}}\\

\medskip
The inclusive $\phi$, $\Lambda$, $K^*$ and
$K_s^0$ signals and their fitted masses from the newly collected $22 \times
10^6$ $J/\psi$ events sample with BESII are shown in Fig. 3.

\begin{figure}[thb]
\centerline{\hbox{\psfig{file=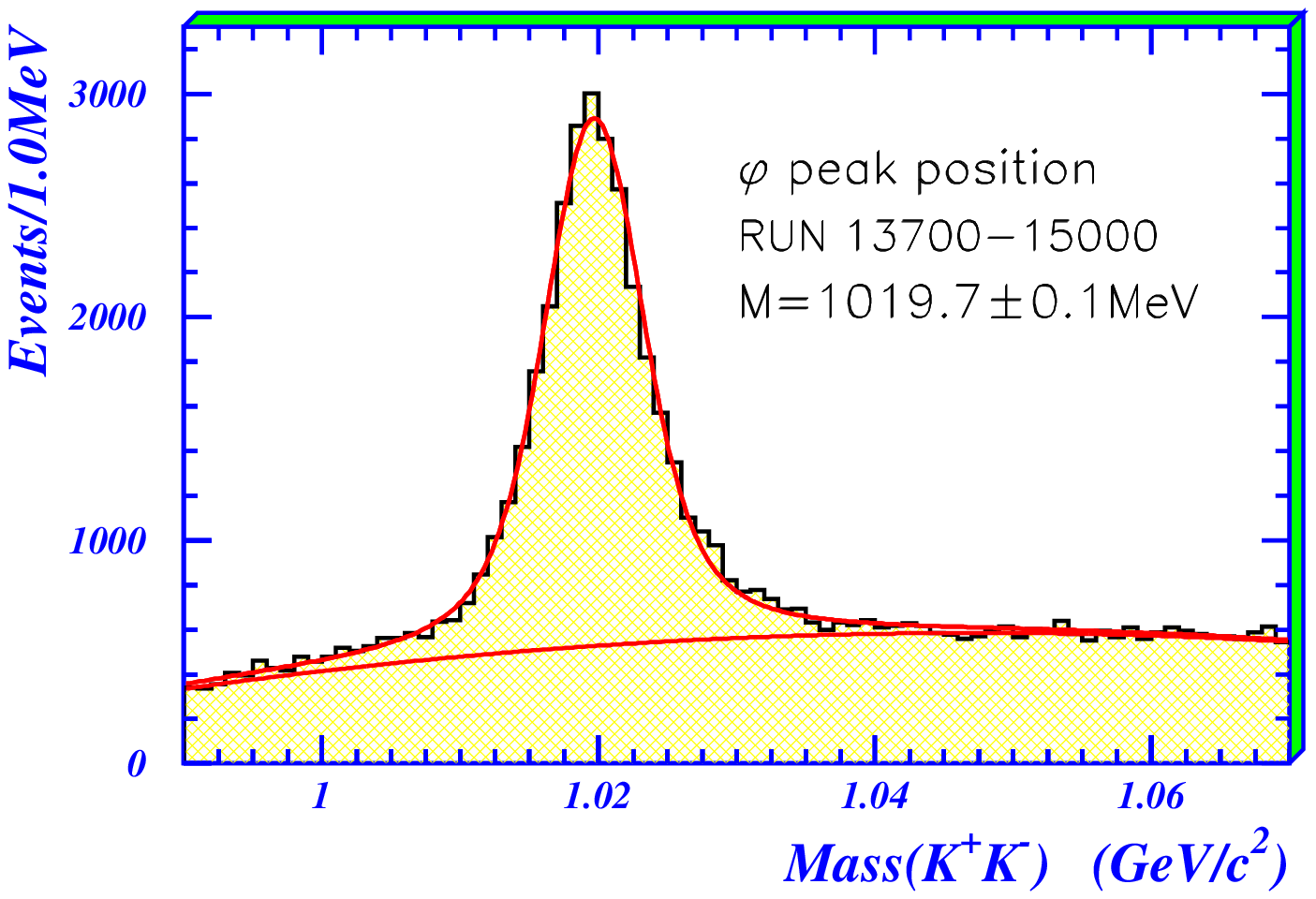,height=30mm,width=32mm}
\psfig{file=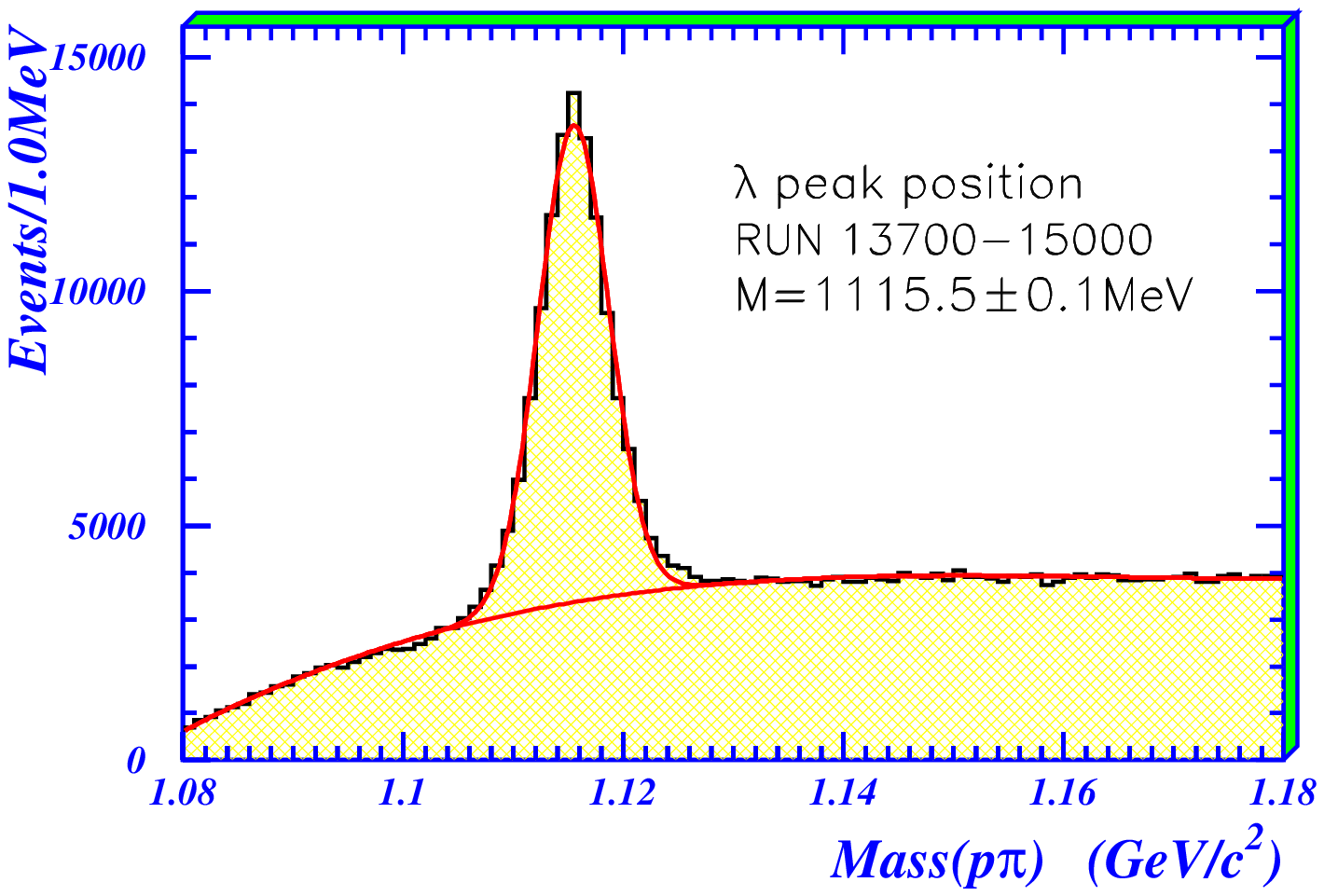,height=30mm,width=32mm}}}
\centerline{\hbox{\psfig{file=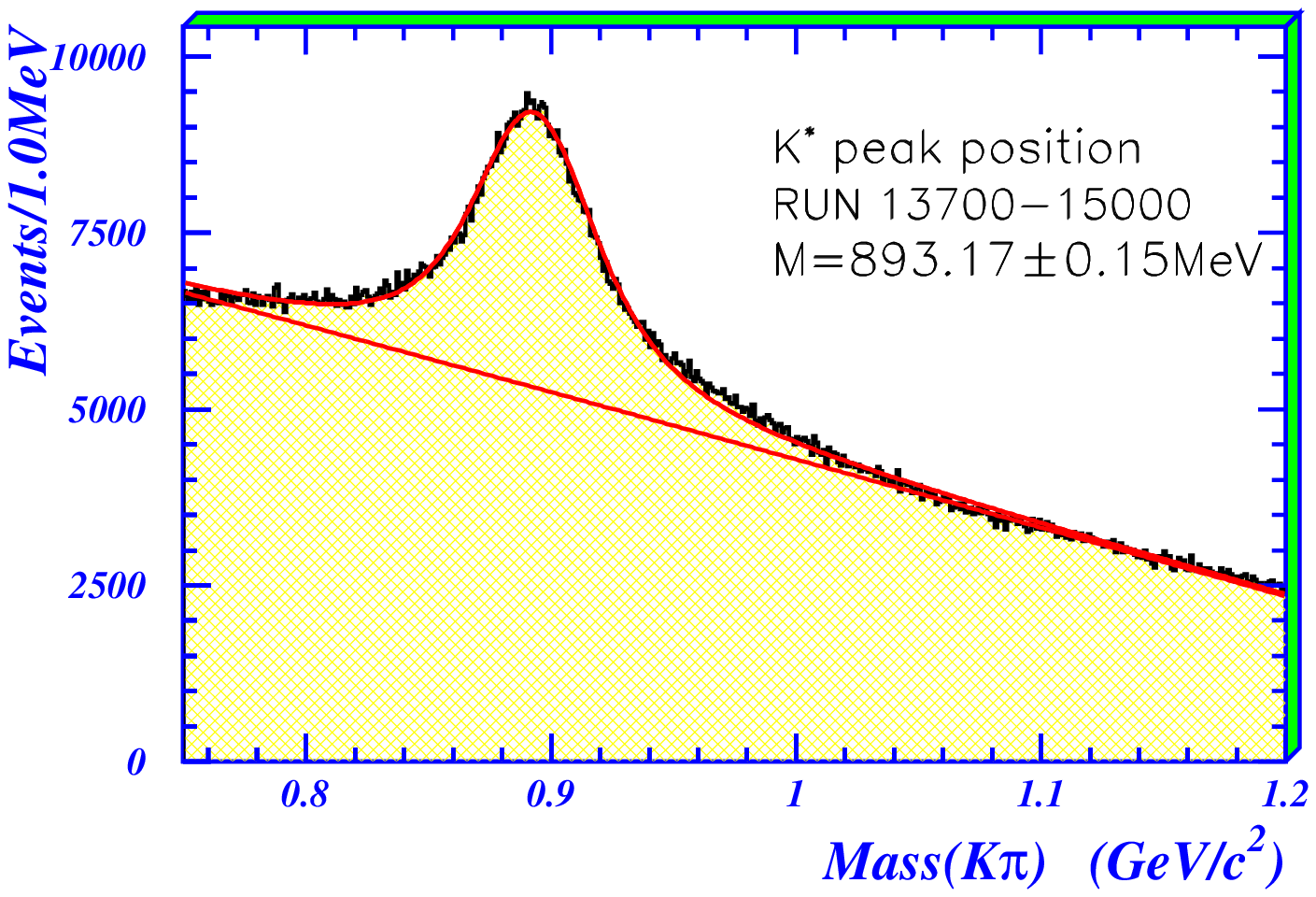,height=30mm,width=32mm}
\psfig{file=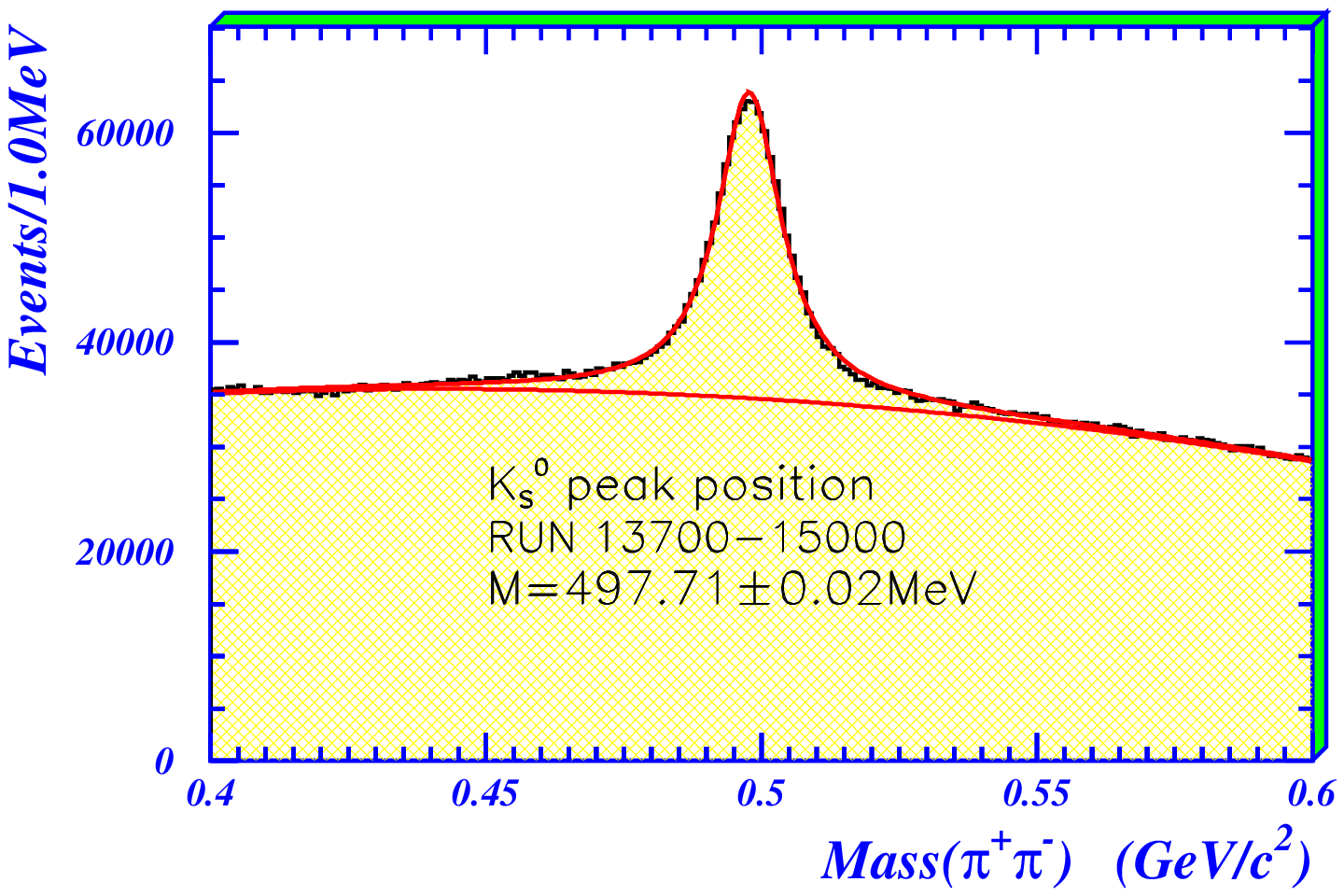,height=30mm,width=32mm}}}
\vspace*{-0.3cm}
\caption{Inclusive $\phi$, $\Lambda$, $K^*$ and $K_s^0$ signals}
\end{figure}

Fig. 4 is the invariant mass spectra and Daliz plot for
$J/\psi \to \omega \pi^+ \pi^-$ decay. Clear $\pi^0$
and $\omega$ signals are seen in 2$\gamma$'s and $\pi^+ \pi^- \pi^0$
invariant mass spectra, respectively. In $\pi^+ \pi^-$ mass distribution,
which recoils against $\omega$, an $f_2(1270)$
and a big bump around 500 MeV
are observed. The invariant masses of $M_{\pi^+ \pi^-}$ and
$M_{K \pi}$ are plotted in Fig. 5 for $J/\psi \to K^{*\pm}K^{\mp}, K^{*\pm}
\to K_s^0 \pi^{\pm}$ and $K_s^0 \to \pi^+ \pi^-$ decay. Both $K_s^0$ and
$K^*$ are nicely peaked.
\begin{figure}[hbt]
\vspace{9pt}
\centerline{\hbox{\psfig{file=wf2_13700-16835.epsi,height=50mm,width=65mm}}}
\caption{$J/\psi \to \omega \pi^+ \pi^-$, $\omega \to \pi^+ \pi^- \pi^0$}
\label{fig4:wpipi}
\end{figure}

\begin{figure}[hbt]
\centerline{\hbox{\psfig{file=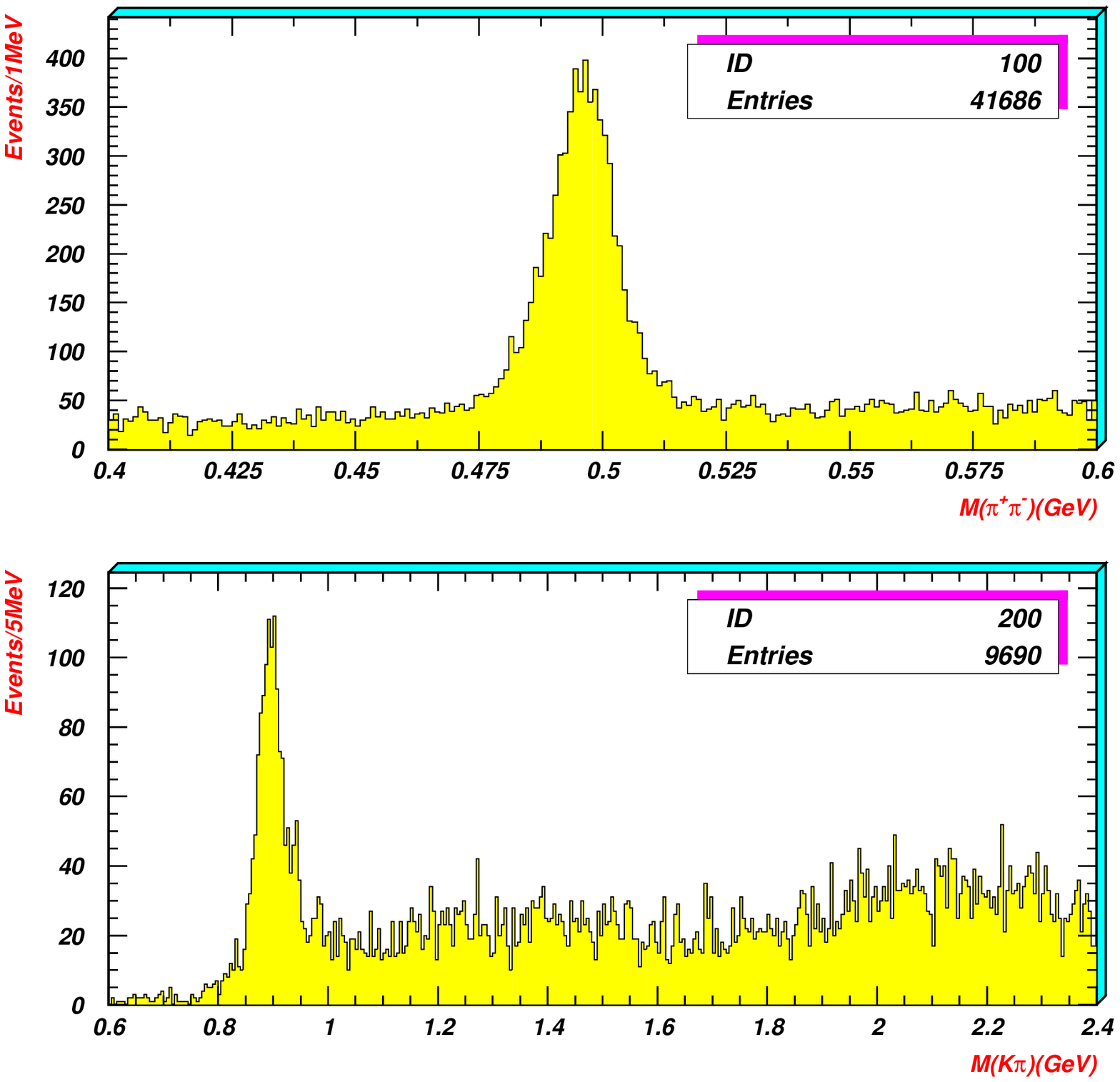,height=50mm,width=50mm}}}
\vspace*{-0.3cm}
\caption{$J/\psi \to K^{*\pm}K^{\mp}$, $K^{*\pm} \to K_s^0 \pi^{\pm}$,
$K_s^0 \to \pi^+ \pi^-$}
\label{fig5:Kstark}
\end{figure}

\begin{figure}[htb]
\vspace{9pt}
\centerline{\hbox{\psfig{file=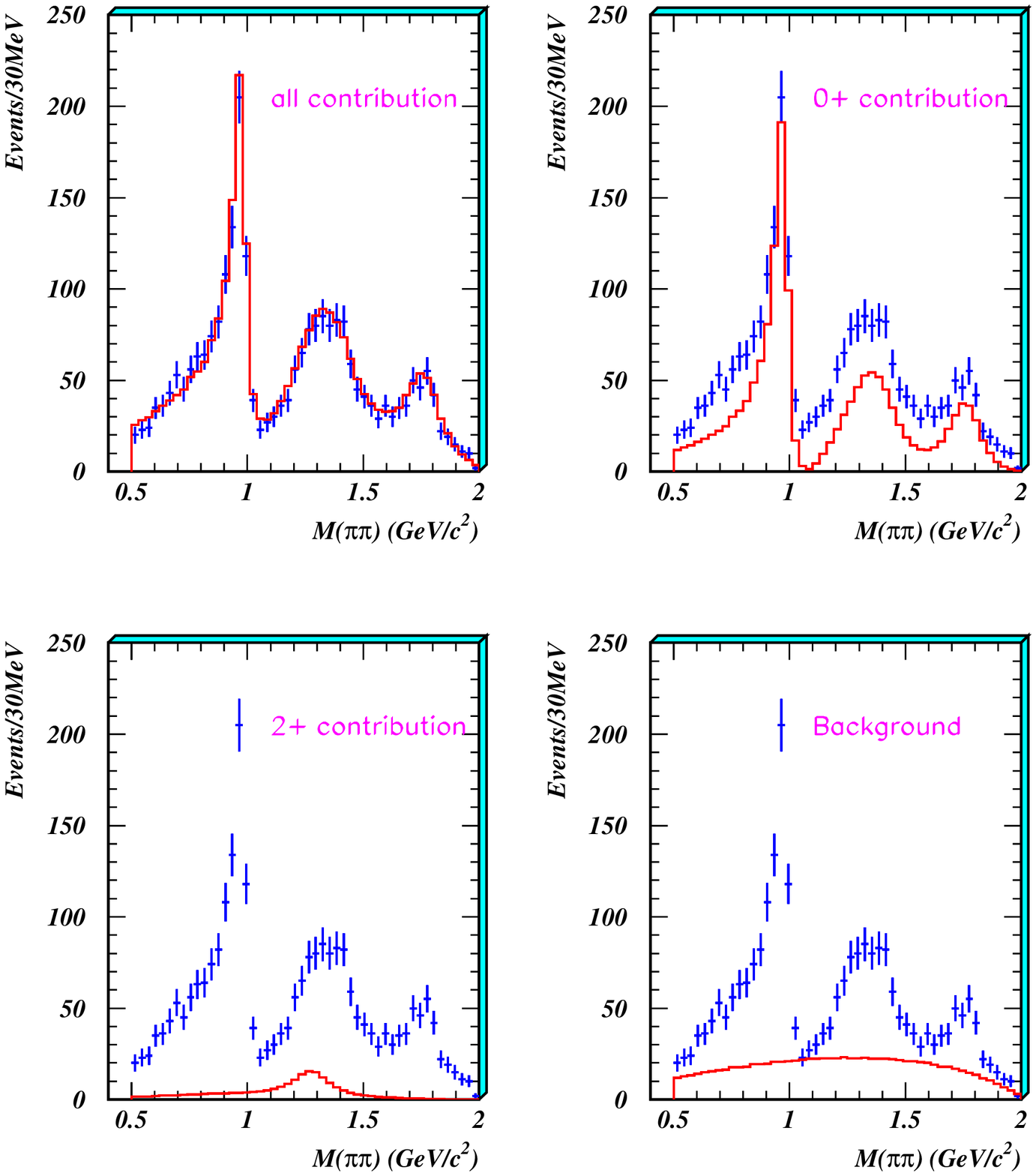,height=55mm,width=75mm}}}
\vspace*{-0.3cm}
\caption{Projection of every component on $\pi^+ \pi^-$ mass in $J/\psi \to
\phi \pi^+ \pi^-$ (Very preliminary).}
\label{fig7:phipipi}
\end{figure}   

\begin{figure}[ht]
\vspace{9pt}
\centerline{\hbox{\psfig{file=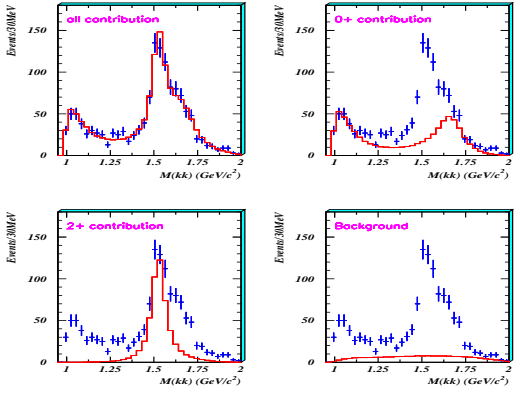,height=55mm,width=75mm}}}
\vspace*{-0.3cm}
\caption{Projection of every component on $K^+ K^-$ mass in $J/\psi \to \phi
K^+ K^-$ (Very preliminary).}
\label{fig8:phikk}
\end{figure}   
   
\bigskip
\noindent $\bullet$ Partial Wave Analyses on  $J/\psi \to \phi \pi^+ \pi^-$
and $\phi K^+ K^-$

\medskip
Partial Wave Analyses of $J/\psi \to \phi \pi^+ \pi^-$
and $\phi K^+ K^-$ are performed.
Fig. 6 represents the contribution of every component from the fit in
$J/\psi \to \phi \pi^+ \pi^-$. Three $0^{++}$, located at 980 MeV,
1370 MeV and 1770 MeV, and one $2^{++}$ at 1270 MeV are observed
in $\pi^+ \pi^-$ invariant mass recoiling against $\phi$.
In $J/\psi \to \phi K^+ K^-$, $f_2'(1525)$ and $f_0(1710)$ components
are found to be needed for a best fit. The projection of different
components on $K^+ K^-$ mass is shown in Fig. 7.

\bigskip
\noindent $\bullet$ Inclusive $\gamma$ spectrum

\medskip
In addition to study the radiative decays
exclusively, the inclusive $\gamma$ spectrum is another place to
search for glueballs. Due to the relatively poor energy
resolution for BSC, we use $\gamma$ conversion to
$e^+ e^-$ pair inside our detector and then measure the momenta of $e^+$
and $e^-$ in MDC, which has a momentum resolution of $1.8\% \sqrt{1+p^2}$
($p$ in GeV).  
The inclusive $\gamma$ spectrum is plotted in Fig. 8.


\begin{figure}[htb]
\centerline{\hbox{\psfig{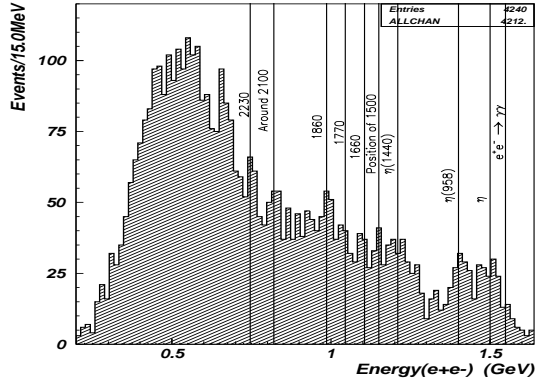}}}
\vspace*{-0.3cm}
\caption{Inclusive $\gamma$ ($E_{e^+e^-}$) spectrum (Very preliminary).}
\label{fig9:inclusivel_g}
\end{figure}

\subsection{Results from $\psi(2S)$ data}

One expects the following relation holds for any hadronic final 
state, according to perturbative QCD,

\begin{eqnarray*}
Q_h \equiv \frac{B(\psi(2S)\rightarrow h)}{B(J/\psi\rightarrow h)}\simeq
    \frac{B(\psi(2S)\rightarrow e^+e^-)}{B(J/\psi\rightarrow e^+e^-)}
  \nonumber \\
    = (14.8 \pm 2.2) \%. 
\end{eqnarray*}

\noindent 
where the leptonic branching fractions are taken from PDG2000~\cite{pdg2k}. 
This relation is referred to  as PQCD $15\%$ rule.

BES measured, for the first time, branching fractions of 12 $\psi(2S)$ hadronic
decays as shown in table 3. 

\begin{table}
\caption{ BES preliminary $B(\psi(2S)\rightarrow h)$ and 15\% rule test
 (limit at C.L.90\%)}
\label{tab:smtab}
\begin{tabular}{|c|c|c|}

\hline

\raisebox{0pt}[12pt][6pt]{$h$} &

\raisebox{0pt}[12pt][6pt]{$B_h (10^{-5})$} &

\raisebox{0pt}[12pt][6pt]{$Q_h (\%)$} \\\hline

$\omega K^+K^-$         &  $12.5\pm 5.6$  &  $16.9\pm 9.4$ \\\hline

$\omega p\overline{p}$  &  $6.4\pm 2.6$   &  $5.0\pm 2.2$  \\\hline

$\phi\pi^+\pi^-$        &  $16.8\pm 3.2$  &  $21.0\pm 5.1$ \\\hline

$\phi K^+K^-$           &  $5.8\pm 2.2$   &  $7.0\pm 2.9$  \\\hline

$\phi p\overline{p}$    & $0.82\pm 0.52$ & $18.1\pm 12.8$ \\\hline

$\phi f_0$              &  $6.3\pm 1.8$   & $19.6\pm 7.8$  \\\hline

$ K^*K^-\pi^+ +c.c.$    &  $60.4\pm 9.0$  &  ?             \\\hline

$ K^*\overline{K}^*+c.c.$ & $3.92\pm 1.03$ & $13.6\pm 4.9$\\\hline

$K^*\overline{K_2}^*+c.c.$ & $7.98\pm 5.28$ &$1.20\pm 0.93$ \\\hline

$\pi^0\pi^+\pi^-p\overline{p}$& $34.9\pm 6.4$ & $15.2\pm 6.6$ \\\hline

$\eta\pi^+\pi^-p\overline{p}$ & $24.7\pm 9.6$   &  ? \\\hline
$\eta p\overline{p}$      & $<18. $             & $ <8.6$   \\\hline
\end{tabular}
\end{table}

     In the context of flavor SU(3) symmetry, a pure $c\overline{c}$ 
state is a
flavor singlet. In the limit of SU(3) flavor symmetry, the  
phase-space-corrected reduced branching ratio $\mid M_i \mid^2$ 

\begin{displaymath}
  \mid M_i \mid^2  =  \frac{B(\psi(2S) \rightarrow B_i\overline{B_i})}
       {\pi p^*/\sqrt{s}}
\end{displaymath}
\noindent
 should be the same for every baryon
 $B_i$ in the same multiplet ($p^*$ is the momentum of the baryon in the 
$\psi(2S)$ rest frame).
 This relation works reasonably well for 
$J/\psi\rightarrow B\overline{B}$, but has not been tested for 
the $\psi(2S)$.

  BES measured, as listed in Table 4, branching fractions of eight 
$\psi(2S) \rightarrow B \overline {B}$ channels, five of which are for 
the first time. 

\begin{table}
\caption{Branching fractions of $\psi(2S)\rightarrow B\overline{B}$
 (limit at C.L.90\%)}
\label{tab:smtab}
{\footnotesize
\begin{tabular}{|c|c|c|c|}

\hline
                       
{ Decay}& {$B$} & { $Q_h$} &$\mid M_i \mid^2_{\psi(2S)}$      \\
&$(10^{-4})$ &$(\%)$ &  $(10^{-4})$ \\ \hline
$p\overline{p}$           &$2.16 \pm .39$  & $10.1 \pm 1.9$ & $1.60\pm
.29$\\
$\Lambda\overline{\Lambda}$    &$1.81 \pm .34$    & $13.4 \pm 2.9$ &
$1.45\pm
.27$\\
$\Sigma^0\overline{\Sigma}^0$  &$ 1.2 \pm .6$       & $9.4 \pm 4.6$& $1.0\pm
.5$ \\
$\Xi^-\overline{\Xi}^+$        & $0.94 \pm .31$   & $10.4 \pm 4.1$& $0.86\pm
.28$ \\\hline
$\Delta^{++}\overline{\Delta}^{--}$ & $1.28 \pm .35$   & $11.6 \pm 4.5$ &
$1.10\pm .30$\\
$\Sigma^{*-}\overline{\Sigma}^{*+}$   & $1.1 \pm .4$   & $ 11 \pm 4$ &
$1.1\pm
 .4$\\
$\Xi^{*0}\overline{\Xi}^{*0}$       &$<.81 $   &  & $<0.93$             \\
$\Omega^-\overline{\Omega}^+$       &$<.73$    &   & $<1.11$
\\\hline

\end{tabular}}
\end{table}     

   The $\mid M_i \mid^2$ for $\psi(2S)\rightarrow B\overline{B}$
have been calculated based on BES data and also listed in Table 4. 
The results show a tendency of smaller values for the 
higher masses, only marginally consistent to flavor-SU(3) symmetry
prediction, similar to $J/\psi$ case.

   By studying the radiative decays of $\psi(2S)\rightarrow\gamma(\pi\pi,
K\overline{K},\eta\eta)$,
BES has measured 8 branching fractions for the first 
time, and their preliminary results are listed in Table 5. The branching 
fractions for $\psi(2S)\rightarrow\gamma f_2(1270)$ and
$\psi(2S)\rightarrow
\gamma f_J(1710)\rightarrow\gamma K\overline{K}$ agree with PQCD $15\%$
 prediction.
The ratio of $\chi_{c0}\rightarrow \eta\eta$ to  $\chi_{c0}\rightarrow
\pi^0\pi^0$ is equal to $0.73\pm 0.39$, consistent with the 
theoretical prediction of 0.95 based on the flavor SU(3) symmetry.

\begin{table}
\caption{Preliminary branching fractions
 from $\psi(2S)$ radiative decay
(limit at C.L.90\%)}
\label{tab:smtab}
\begin{tabular}{|c|c|}

\hline

\raisebox{0pt}[12pt][6pt]{Process} &

\raisebox{0pt}[12pt][6pt]{$B (10^{-4})$} \\\hline

$\psi(2S)\rightarrow\gamma f_2(1270)$  &
            $2.27\pm 0.43$\\ \hline

$\rightarrow\gamma f_J(1710)\rightarrow\gamma\pi\pi$
  & $.336\pm .165$   \\ \hline

$\rightarrow\gamma f_J(1710)\rightarrow\gamma K^+K^-$
  & $0.55\pm 0.21$   \\ \hline

$\rightarrow\gamma f_J(1710)\rightarrow\gamma K^0_S K^0_S$
  & $0.21\pm 0.15$   \\ \hline

$\chi_{c0}\rightarrow\pi^0\pi^0$  & $26.8\pm 6.5$  \\ \hline

$\chi_{c2}\rightarrow\pi^0\pi^0$  & $8.8\pm 5.6$ \\ \hline

$\chi_{c0}\rightarrow\eta\eta$  & $19.4\pm 10.0$ \\ \hline

$\chi_{c2}\rightarrow\eta\eta$  & $ < 12.2$  \\ \hline
\end{tabular}
\end{table}

\section{BES future plans}

The plan for our near future is to collect about $50 \times 
10^6 $ $J/\psi$ events with BESII before 2001. With this 
amount of $J/\psi$ event sample, about 6 times as more data as the
largest sample existing in the world, BES collaboration can systematically 
study the light hadron spectroscopy and $N^*$
baryons; search for glueball and hybrid states; as well as exploring the rare
decays in $J/\psi$ decays.       

The plan for the middle term future is to significantly upgrade 
BEPC and BESII. The upgraded machine and detector will be named BEPCII and 
BESIII. The upgrade goal for the machine
is to increase the peak luminosity from about $5 \times 10^{30} /cm^2 s$ to
about $5 \times 10^{31} /cm^2 s$ at the $J/\psi$ resonance. Multi-bunch
train will be adopted and the bunch length will be shorten
to about 1 cm. In addition, the injection rate will be significantly increased. For
the detector, we will build an new electromagnetic calorimeter using
scintilating fiber with lead, which gives an energy resolution better than
10\%/$\sqrt{E}$. The time-of-flight may also be rebuilt to have a resolution
less than 150 ps. We are also considering building a drift chamber using 
Helium based gas and Aluminum field wires. New vertex chamber and luminosity 
monitor will be built to adapt the new interaction region.  

The upgrade project has been endorsed by the Chinese central government with
a budget about 40 M US dollars. R\&D program has been started for both
machine and detector.

\end{document}